\documentclass[prl,preprint,amsfonts,showpacs,showkeys]{revtex4}
\usepackage{graphicx}
\RequirePackage{times}
\RequirePackage{mathptm}

\begin{document}
%\sloppy

\title{Microphysical analogues of flyby anomalies}

\author{Karl Svozil}
\email{svozil@tuwien.ac.at}
\homepage{http://tph.tuwien.ac.at/~svozil}
\affiliation{Institute for Theoretical Physics, Vienna University of Technology,
Wiedner Hauptstra\ss e 8-10/136, A-1040 Vienna, Austria}

\begin{abstract}
We discuss Doppler shift and interferometric measurements in analogy to the recently reported macroscopic flyby anomalies.
\end{abstract}


\maketitle

The recently reported anomalous orbital-energy changes observed during six Earth flybys by the
Galileo, NEAR, Cassini, Rosetta, and MESSENGER spacecraft documented an unexpected frequency increase
in the post encounter radio Doppler data generated by stations of the NASA Deep Space Network \cite{anderson:091102,anderson:newast}.
These observations deserve great attention, mainly because of the possible consequences for theories of gravitation and electromagnetism~\cite{Dittus}
and of other issues in astrophysics~\cite{adler-arXiv:0805.2895,cahill-2008}.

In what follows we shall investigate microscopic analogues of macroscopic flyby configurations for which anomalies have been noted.
We first start with a straightforward discussion of flyby configurations and later concentrate on interferometric setups.

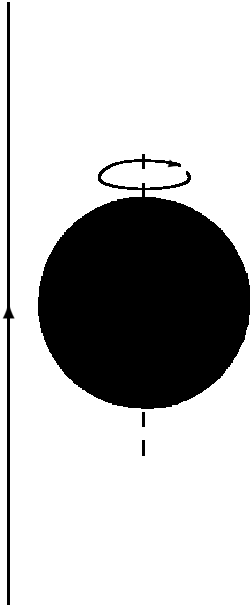
\begin{figure}
\centering
%TeXCAD Picture [1.pic]. Options:
%\grade{\on}
%\emlines{\off}
%\epic{\off}
%\beziermacro{\on}
%\reduce{\on}
%\snapping{\off}
%\pvinsert{% Your \input, \def, etc. here}
%\quality{8.000}
%\graddiff{0.005}
%\snapasp{1}
%\zoom{26.9087}
\unitlength 2mm % = 2.845pt
\linethickness{0.5pt}
\ifx\plotpoint\undefined\newsavebox{\plotpoint}\fi % GNUPLOT compatibility
\begin{picture}(17,40)(0,0)
%\end
\put(10,20){\circle{14}}
%\dashline{1}(10,10)(10,30)
\put(9.93,9.93){\line(0,1){.9524}}
\put(9.93,11.834){\line(0,1){.9524}}
\put(9.93,13.739){\line(0,1){.9524}}
\put(9.93,15.644){\line(0,1){.9524}}
\put(9.93,17.549){\line(0,1){.9524}}
\put(9.93,19.454){\line(0,1){.9524}}
\put(9.93,21.358){\line(0,1){.9524}}
\put(9.93,23.263){\line(0,1){.9524}}
\put(9.93,25.168){\line(0,1){.9524}}
\put(9.93,27.073){\line(0,1){.9524}}
\put(9.93,28.977){\line(0,1){.9524}}
%\end
\qbezier(7.172,28.727)(6.281,27.723)(9.997,27.612)
\qbezier(12.821,28.727)(13.713,27.723)(9.997,27.612)
\qbezier(9.997,29.507)(7.916,29.526)(7.172,28.727)
\qbezier(12.338,29.173)(11.948,29.451)(9.997,29.507)
%\vector(11,29.47)(12.264,29.21)
\put(12.264,29.21){\vector(4,-1){.07}}\multiput(11,29.47)(.157941,-.032517){8}{\line(1,0){.157941}}
%\end
%\circle*(9.997,20.031){13.919}
\put(5.0195,15.0535){\rule{9.9545\unitlength}{9.9545\unitlength}}
\multiput(7.2773,24.8955)(0,-11.3506){2}{\rule{5.4389\unitlength}{1.6211\unitlength}}
\multiput(8.5828,26.4041)(0,-13.2552){2}{\rule{2.8279\unitlength}{.5085\unitlength}}
\multiput(9.2584,26.8001)(0,-13.7515){2}{\rule{1.4767\unitlength}{.2127\unitlength}}
\multiput(10.6227,26.8001)(-1.7033,0){2}{\multiput(0,0)(0,-13.7097){2}{\rule{.4515\unitlength}{.1709\unitlength}}}
\multiput(11.2982,26.4041)(-3.3779,0){2}{\multiput(0,0)(0,-13.0893){2}{\rule{.775\unitlength}{.3425\unitlength}}}
\multiput(11.2982,26.6342)(-3.0487,0){2}{\multiput(0,0)(0,-13.4105){2}{\rule{.4457\unitlength}{.2036\unitlength}}}
\multiput(11.9607,26.4041)(-4.3647,0){2}{\multiput(0,0)(0,-12.9821){2}{\rule{.4368\unitlength}{.2354\unitlength}}}
\multiput(12.2851,26.4041)(-4.8492,0){2}{\multiput(0,0)(0,-12.9226){2}{\rule{.2726\unitlength}{.1759\unitlength}}}
\multiput(12.6038,24.8955)(-6.5296,0){2}{\multiput(0,0)(0,-10.7075){2}{\rule{1.3156\unitlength}{.978\unitlength}}}
\multiput(12.6038,25.761)(-5.9438,0){2}{\multiput(0,0)(0,-11.9241){2}{\rule{.7299\unitlength}{.4636\unitlength}}}
\multiput(12.6038,26.1121)(-5.6387,0){2}{\multiput(0,0)(0,-12.4288){2}{\rule{.4247\unitlength}{.266\unitlength}}}
\multiput(12.6038,26.2656)(-5.4834,0){2}{\multiput(0,0)(0,-12.6535){2}{\rule{.2695\unitlength}{.1836\unitlength}}}
\multiput(12.916,26.1121)(-6.1045,0){2}{\multiput(0,0)(0,-12.3539){2}{\rule{.266\unitlength}{.1911\unitlength}}}
\multiput(13.2212,25.761)(-6.8584,0){2}{\multiput(0,0)(0,-11.7557){2}{\rule{.4097\unitlength}{.2952\unitlength}}}
\multiput(13.2212,25.9437)(-6.7109,0){2}{\multiput(0,0)(0,-12.0245){2}{\rule{.2621\unitlength}{.1985\unitlength}}}
\multiput(13.5184,25.761)(-7.3011,0){2}{\multiput(0,0)(0,-11.6661){2}{\rule{.2579\unitlength}{.2056\unitlength}}}
\multiput(13.8069,24.8955)(-8.2814,0){2}{\multiput(0,0)(0,-10.3007){2}{\rule{.661\unitlength}{.5711\unitlength}}}
\multiput(13.8069,25.3541)(-8.0121,0){2}{\multiput(0,0)(0,-10.9695){2}{\rule{.3917\unitlength}{.3226\unitlength}}}
\multiput(13.8069,25.5643)(-7.8737,0){2}{\multiput(0,0)(0,-11.2797){2}{\rule{.2533\unitlength}{.2125\unitlength}}}
\multiput(14.0862,25.3541)(-8.4273,0){2}{\multiput(0,0)(0,-10.866){2}{\rule{.2484\unitlength}{.2192\unitlength}}}
\multiput(14.3555,24.8955)(-9.0886,0){2}{\multiput(0,0)(0,-10.0775){2}{\rule{.3711\unitlength}{.348\unitlength}}}
\multiput(14.3555,25.131)(-8.9606,0){2}{\multiput(0,0)(0,-10.4262){2}{\rule{.2432\unitlength}{.2256\unitlength}}}
\multiput(14.4862,25.131)(-9.1556,0){2}{\multiput(0,0)(0,-10.37){2}{\rule{.1768\unitlength}{.1694\unitlength}}}
\multiput(14.6142,24.8955)(-9.4724,0){2}{\multiput(0,0)(0,-9.9613){2}{\rule{.2376\unitlength}{.2318\unitlength}}}
\multiput(14.6142,25.0148)(-9.4102,0){2}{\multiput(0,0)(0,-10.1391){2}{\rule{.1754\unitlength}{.171\unitlength}}}
\multiput(14.7393,24.8955)(-9.659,0){2}{\multiput(0,0)(0,-9.902){2}{\rule{.174\unitlength}{.1725\unitlength}}}
\multiput(14.8615,17.3112)(-11.3506,0){2}{\rule{1.6211\unitlength}{5.4389\unitlength}}
\multiput(14.8615,22.6377)(-10.7075,0){2}{\multiput(0,0)(0,-6.5296){2}{\rule{.978\unitlength}{1.3156\unitlength}}}
\multiput(14.8615,23.8409)(-10.3007,0){2}{\multiput(0,0)(0,-8.2814){2}{\rule{.5711\unitlength}{.661\unitlength}}}
\multiput(14.8615,24.3894)(-10.0775,0){2}{\multiput(0,0)(0,-9.0886){2}{\rule{.348\unitlength}{.3711\unitlength}}}
\multiput(14.8615,24.6481)(-9.9613,0){2}{\multiput(0,0)(0,-9.4724){2}{\rule{.2318\unitlength}{.2376\unitlength}}}
\multiput(14.8615,24.7732)(-9.902,0){2}{\multiput(0,0)(0,-9.659){2}{\rule{.1725\unitlength}{.174\unitlength}}}
\multiput(14.9808,24.6481)(-10.1391,0){2}{\multiput(0,0)(0,-9.4102){2}{\rule{.171\unitlength}{.1754\unitlength}}}
\multiput(15.0971,24.3894)(-10.4262,0){2}{\multiput(0,0)(0,-8.9606){2}{\rule{.2256\unitlength}{.2432\unitlength}}}
\multiput(15.0971,24.5201)(-10.37,0){2}{\multiput(0,0)(0,-9.1556){2}{\rule{.1694\unitlength}{.1768\unitlength}}}
\multiput(15.3202,23.8409)(-10.9695,0){2}{\multiput(0,0)(0,-8.0121){2}{\rule{.3226\unitlength}{.3917\unitlength}}}
\multiput(15.3202,24.1201)(-10.866,0){2}{\multiput(0,0)(0,-8.4273){2}{\rule{.2192\unitlength}{.2484\unitlength}}}
\multiput(15.5303,23.8409)(-11.2797,0){2}{\multiput(0,0)(0,-7.8737){2}{\rule{.2125\unitlength}{.2533\unitlength}}}
\multiput(15.727,22.6377)(-11.9241,0){2}{\multiput(0,0)(0,-5.9438){2}{\rule{.4636\unitlength}{.7299\unitlength}}}
\multiput(15.727,23.2551)(-11.7557,0){2}{\multiput(0,0)(0,-6.8584){2}{\rule{.2952\unitlength}{.4097\unitlength}}}
\multiput(15.727,23.5523)(-11.6661,0){2}{\multiput(0,0)(0,-7.3011){2}{\rule{.2056\unitlength}{.2579\unitlength}}}
\multiput(15.9098,23.2551)(-12.0245,0){2}{\multiput(0,0)(0,-6.7109){2}{\rule{.1985\unitlength}{.2621\unitlength}}}
\multiput(16.0781,22.6377)(-12.4288,0){2}{\multiput(0,0)(0,-5.6387){2}{\rule{.266\unitlength}{.4247\unitlength}}}
\multiput(16.0781,22.95)(-12.3539,0){2}{\multiput(0,0)(0,-6.1045){2}{\rule{.1911\unitlength}{.266\unitlength}}}
\multiput(16.2317,22.6377)(-12.6535,0){2}{\multiput(0,0)(0,-5.4834){2}{\rule{.1836\unitlength}{.2695\unitlength}}}
\multiput(16.3701,18.6168)(-13.2552,0){2}{\rule{.5085\unitlength}{2.8279\unitlength}}
\multiput(16.3701,21.3322)(-13.0893,0){2}{\multiput(0,0)(0,-3.3779){2}{\rule{.3425\unitlength}{.775\unitlength}}}
\multiput(16.3701,21.9947)(-12.9821,0){2}{\multiput(0,0)(0,-4.3647){2}{\rule{.2354\unitlength}{.4368\unitlength}}}
\multiput(16.3701,22.319)(-12.9226,0){2}{\multiput(0,0)(0,-4.8492){2}{\rule{.1759\unitlength}{.2726\unitlength}}}
\multiput(16.6002,21.3322)(-13.4105,0){2}{\multiput(0,0)(0,-3.0487){2}{\rule{.2036\unitlength}{.4457\unitlength}}}
\multiput(16.7662,19.2923)(-13.7515,0){2}{\rule{.2127\unitlength}{1.4767\unitlength}}
\multiput(16.7662,20.6566)(-13.7097,0){2}{\multiput(0,0)(0,-1.7033){2}{\rule{.1709\unitlength}{.4515\unitlength}}}
\put(16.956,20.031){\line(0,1){.3976}}
\put(16.945,20.428){\line(0,1){.3963}}
\put(16.911,20.825){\line(0,1){.3937}}
\put(16.854,21.218){\line(0,1){.3898}}
\put(16.775,21.608){\line(0,1){.3847}}
\multiput(16.674,21.993)(-.03075,.09457){4}{\line(0,1){.09457}}
\multiput(16.551,22.371)(-.028882,.074123){5}{\line(0,1){.074123}}
\multiput(16.406,22.742)(-.03307,.072352){5}{\line(0,1){.072352}}
\multiput(16.241,23.103)(-.030958,.05862){6}{\line(0,1){.05862}}
\multiput(16.055,23.455)(-.029363,.048648){7}{\line(0,1){.048648}}
\multiput(15.85,23.796)(-.032094,.046891){7}{\line(0,1){.046891}}
\multiput(15.625,24.124)(-.03038,.039358){8}{\line(0,1){.039358}}
\multiput(15.382,24.439)(-.032579,.037558){8}{\line(0,1){.037558}}
\multiput(15.121,24.739)(-.030819,.031676){9}{\line(0,1){.031676}}
\multiput(14.844,25.024)(-.036651,.033597){8}{\line(-1,0){.036651}}
\multiput(14.551,25.293)(-.038511,.031448){8}{\line(-1,0){.038511}}
\multiput(14.243,25.545)(-.045994,.033367){7}{\line(-1,0){.045994}}
\multiput(13.921,25.778)(-.047825,.030685){7}{\line(-1,0){.047825}}
\multiput(13.586,25.993)(-.05775,.032553){6}{\line(-1,0){.05775}}
\multiput(13.24,26.188)(-.059515,.029201){6}{\line(-1,0){.059515}}
\multiput(12.882,26.364)(-.073304,.030904){5}{\line(-1,0){.073304}}
\multiput(12.516,26.518)(-.09369,.03333){4}{\line(-1,0){.09369}}
\put(12.141,26.651){\line(-1,0){.3818}}
\put(11.759,26.763){\line(-1,0){.3875}}
\put(11.372,26.853){\line(-1,0){.392}}
\put(10.98,26.92){\line(-1,0){.3952}}
\put(10.585,26.965){\line(-1,0){.3971}}
\put(10.188,26.987){\line(-1,0){.3978}}
\put(9.79,26.987){\line(-1,0){.3971}}
\put(9.393,26.964){\line(-1,0){.3951}}
\put(8.998,26.918){\line(-1,0){.3918}}
\put(8.606,26.85){\line(-1,0){.3873}}
\put(8.218,26.759){\line(-1,0){.3815}}
\multiput(7.837,26.646)(-.09361,-.03355){4}{\line(-1,0){.09361}}
\multiput(7.463,26.512)(-.073232,-.031074){5}{\line(-1,0){.073232}}
\multiput(7.096,26.357)(-.059447,-.029339){6}{\line(-1,0){.059447}}
\multiput(6.74,26.181)(-.057674,-.032687){6}{\line(-1,0){.057674}}
\multiput(6.394,25.985)(-.047753,-.030796){7}{\line(-1,0){.047753}}
\multiput(6.059,25.769)(-.045916,-.033474){7}{\line(-1,0){.045916}}
\multiput(5.738,25.535)(-.038437,-.031537){8}{\line(-1,0){.038437}}
\multiput(5.43,25.283)(-.036573,-.033682){8}{\line(-1,0){.036573}}
\multiput(5.138,25.013)(-.030746,-.031748){9}{\line(0,-1){.031748}}
\multiput(4.861,24.727)(-.032492,-.037634){8}{\line(0,-1){.037634}}
\multiput(4.601,24.426)(-.030289,-.039429){8}{\line(0,-1){.039429}}
\multiput(4.359,24.111)(-.031985,-.046966){7}{\line(0,-1){.046966}}
\multiput(4.135,23.782)(-.029249,-.048716){7}{\line(0,-1){.048716}}
\multiput(3.93,23.441)(-.030821,-.058692){6}{\line(0,-1){.058692}}
\multiput(3.745,23.089)(-.032902,-.072429){5}{\line(0,-1){.072429}}
\multiput(3.581,22.727)(-.02871,-.07419){5}{\line(0,-1){.07419}}
\multiput(3.437,22.356)(-.03053,-.09464){4}{\line(0,-1){.09464}}
\put(3.315,21.977){\line(0,-1){.3849}}
\put(3.215,21.592){\line(0,-1){.39}}
\put(3.137,21.202){\line(0,-1){.3938}}
\put(3.081,20.809){\line(0,-1){.3964}}
\put(3.048,20.412){\line(0,-1){1.1914}}
\put(3.085,19.221){\line(0,-1){.3936}}
\put(3.142,18.827){\line(0,-1){.3896}}
\put(3.222,18.438){\line(0,-1){.3844}}
\multiput(3.324,18.053)(.03097,-.09449){4}{\line(0,-1){.09449}}
\multiput(3.448,17.675)(.029054,-.074056){5}{\line(0,-1){.074056}}
\multiput(3.593,17.305)(.033238,-.072275){5}{\line(0,-1){.072275}}
\multiput(3.76,16.943)(.031094,-.058548){6}{\line(0,-1){.058548}}
\multiput(3.946,16.592)(.029475,-.04858){7}{\line(0,-1){.04858}}
\multiput(4.153,16.252)(.032203,-.046816){7}{\line(0,-1){.046816}}
\multiput(4.378,15.924)(.030472,-.039288){8}{\line(0,-1){.039288}}
\multiput(4.622,15.61)(.032666,-.037483){8}{\line(0,-1){.037483}}
\multiput(4.883,15.31)(.030893,-.031605){9}{\line(0,-1){.031605}}
\multiput(5.161,15.026)(.036729,-.033511){8}{\line(1,0){.036729}}
\multiput(5.455,14.758)(.038584,-.031358){8}{\line(1,0){.038584}}
\multiput(5.764,14.507)(.046071,-.03326){7}{\line(1,0){.046071}}
\multiput(6.086,14.274)(.047896,-.030574){7}{\line(1,0){.047896}}
\multiput(6.421,14.06)(.057825,-.032419){6}{\line(1,0){.057825}}
\multiput(6.768,13.865)(.059583,-.029062){6}{\line(1,0){.059583}}
\multiput(7.126,13.691)(.073375,-.030733){5}{\line(1,0){.073375}}
\multiput(7.493,13.537)(.09376,-.03311){4}{\line(1,0){.09376}}
\put(7.868,13.405){\line(1,0){.382}}
\put(8.25,13.294){\line(1,0){.3877}}
\put(8.637,13.205){\line(1,0){.3922}}
\put(9.03,13.139){\line(1,0){.3953}}
\put(9.425,13.095){\line(1,0){.3972}}
\put(9.822,13.074){\line(1,0){.3978}}
\put(10.22,13.075){\line(1,0){.397}}
\put(10.617,13.099){\line(1,0){.395}}
\put(11.012,13.146){\line(1,0){.3917}}
\put(11.404,13.215){\line(1,0){.3871}}
\put(11.791,13.307){\line(1,0){.3812}}
\multiput(12.172,13.42)(.074825,.027013){5}{\line(1,0){.074825}}
\multiput(12.546,13.555)(.073159,.031244){5}{\line(1,0){.073159}}
\multiput(12.912,13.711)(.059379,.029477){6}{\line(1,0){.059379}}
\multiput(13.268,13.888)(.057598,.032821){6}{\line(1,0){.057598}}
\multiput(13.614,14.085)(.047682,.030907){7}{\line(1,0){.047682}}
\multiput(13.947,14.301)(.045838,.033581){7}{\line(1,0){.045838}}
\multiput(14.268,14.536)(.038364,.031627){8}{\line(1,0){.038364}}
\multiput(14.575,14.789)(.03244,.030015){9}{\line(1,0){.03244}}
\multiput(14.867,15.06)(.030672,.031819){9}{\line(0,1){.031819}}
\multiput(15.143,15.346)(.032404,.037709){8}{\line(0,1){.037709}}
\multiput(15.402,15.648)(.030197,.039499){8}{\line(0,1){.039499}}
\multiput(15.644,15.964)(.031876,.04704){7}{\line(0,1){.04704}}
\multiput(15.867,16.293)(.029136,.048784){7}{\line(0,1){.048784}}
\multiput(16.071,16.634)(.030685,.058764){6}{\line(0,1){.058764}}
\multiput(16.255,16.987)(.032733,.072505){5}{\line(0,1){.072505}}
\multiput(16.419,17.35)(.028537,.074257){5}{\line(0,1){.074257}}
\multiput(16.562,17.721)(.03031,.09471){4}{\line(0,1){.09471}}
\put(16.683,18.1){\line(0,1){.3851}}
\put(16.782,18.485){\line(0,1){.3902}}
\put(16.859,18.875){\line(0,1){.394}}
\put(16.914,19.269){\line(0,1){.7618}}
%\end
\thicklines
%\vector[middle](1,0)(1,40)
\put(1,20){\vector(0,1){.07}}\put(1,0){\line(0,1){40}}
\end{picture}
\caption{A particle beam flies by a heavy rotating (magnetic) object.
The flyby should be along a path parallel to the rotation axis.}
\label{2008-flyby-f-simple_flyby}
\end{figure}

In a first approach consider a beam of neutral particles, such as neutrons or photons, being deflected by a heavy rotating (magnetic) object.
The flyby should be along a path parallel to the rotation axis.
This configuration is depicted in Fig.~\ref{2008-flyby-f-simple_flyby}.

For the prediction of the frequency shift, we take formula~(2) of Ref.~\cite{anderson:091102}
$$
\Delta \omega = 2K\omega (\cos \delta_i -\cos \delta_o),
$$
with $\delta_i$ and $\delta_o$ as the declinations of the incoming and outgoing osculating asymptotic velocity vectors,
and $K= 2\Omega R /c$, where $\Omega$ is the angular rotational velocity and $R$ the mean radius of the heavy object. $c$ stands for the speed of light in vacuum.
In our case, as $\delta_i =0$ and $\delta_o = \pi$, we obtain as an estimate for the frequency shift
\begin{equation}
\Delta \omega = {8\Omega R\over c}\;\omega .
\end{equation}
One straightforward way to realize heavy (magnetic) rotating objects are stacks of data disks/hard drives which operate at about 7200 revolutions per minute, corresponding to 120~Hz or 754~rad/s.
This is about seven orders of magnitude greater than Earth's angular rotation velocity.
Typical disks have a radius of about 0.05~m, which is about eight orders of magnitude smaller than Earth's mean radius.
Thus, in this case,
\begin{equation}
K= 2.513\times 10^{-7}
\end{equation}
is only about a factor of ten smaller than for Earth.

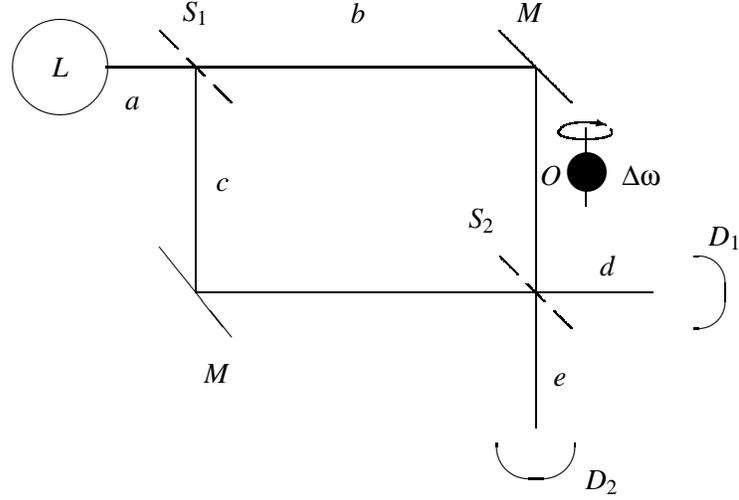
\begin{figure}
\centering

%TeXCAD Picture [1.pic]. Options:
%\grade{\off}
%\emlines{\off}
%\epic{\off}
%\beziermacro{\off}
%\reduce{\on}
%\snapping{\off}
%\quality{0.200}
%\graddiff{0.010}
%\snapasp{1}
%\zoom{11.3137}
\unitlength 1.2mm % = 1.992pt
\linethickness{0.4pt}
\ifx\plotpoint\undefined\newsavebox{\plotpoint}\fi % GNUPLOT compatibility
\begin{picture}(83.67,61)(0,0)
\put(62.67,40){\line(0,-1){25}}
\put(10,55){\makebox(0,0)[cc]{$L$}}
\put(10,55){\circle{10}}
\put(15,55){\line(1,0){40}}
\put(44.67,55){\line(1,0){18}}
\put(25,55){\line(0,-1){25}}
\put(25,30){\line(1,0){13}}
\put(62.67,55){\line(0,-1){25}}
\put(62.67,30){\line(-1,0){35}}
\put(62.67,30){\line(1,0){13}}
\put(62.67,30){\line(0,-1){13}}
%\emline(58.67,59)(66.67,51)
\multiput(58.67,59)(.048192771,-.048192771){166}{\line(0,-1){.048192771}}
%\end
\put(21,35){\line(4,-5){8}}
\put(80.17,30){\oval(7,8)[r]}
\put(83.67,36){\makebox(0,0)[cc]{$D_1$}}
\put(28,42){\makebox(0,0)[cc]{$c$}}
\put(25,61){\makebox(0,0)[cc]{$S_1$}}
\put(56.67,38){\makebox(0,0)[cc]{$S_2$}}
%\emline(24,56)(26,54)
\multiput(24,56)(.04761905,-.04761905){42}{\line(0,-1){.04761905}}
%\end
%\emline(23,57)(21,59)
\multiput(23,57)(-.04761905,.04761905){42}{\line(0,1){.04761905}}
%\end
%\emline(27,53)(29,51)
\multiput(27,53)(.04761905,-.04761905){42}{\line(0,-1){.04761905}}
%\end
%\emline(61.67,31)(63.67,29)
\multiput(61.67,31)(.04761905,-.04761905){42}{\line(0,-1){.04761905}}
%\end
%\emline(60.67,32)(58.67,34)
\multiput(60.67,32)(-.04761905,.04761905){42}{\line(0,1){.04761905}}
%\end
%\emline(64.67,28)(66.67,26)
\multiput(64.67,28)(.04761905,-.04761905){42}{\line(0,-1){.04761905}}
%\end
\put(18,51){\makebox(0,0)[cc]{$a$}}
\put(74.327,43){\makebox(0,0)[cc]{$\Delta \omega$}}
\put(64.327,43){\makebox(0,0)[cc]{$O$}}
\put(70.67,33){\makebox(0,0)[cc]{$d$}}
\put(62.67,13.33){\oval(8.67,8)[b]}
\put(70,9){\makebox(0,0)[cc]{$D_2$}}
\put(65.33,20.33){\makebox(0,0)[cc]{$e$}}
\put(43,61){\makebox(0,0)[cc]{$b$}}
\put(62,61){\makebox(0,0)[cc]{$M$}}
\put(27.33,21){\makebox(0,0)[cc]{$M$}}
\put(68.324,43.31){\circle*{5.127}}
\put(68.236,39.598){\line(0,1){8.662}}
%\qbezier(65.332,48.084)(64.441,47.08)(68.157,46.969)
\put(65.332,48.084){\line(0,-1){.0484}}
\put(65.291,48.036){\line(0,-1){.0474}}
\put(65.256,47.988){\line(0,-1){.0463}}
\put(65.226,47.942){\line(0,-1){.0453}}
\put(65.202,47.897){\line(0,-1){.0442}}
\put(65.183,47.852){\line(0,-1){.2796}}
\put(65.205,47.573){\line(0,-1){.0357}}
\put(65.23,47.537){\line(0,-1){.0346}}
\put(65.26,47.503){\line(1,0){.036}}
\put(65.296,47.469){\line(1,0){.0415}}
\put(65.338,47.436){\line(1,0){.047}}
\put(65.385,47.405){\line(1,0){.0525}}
\put(65.437,47.375){\line(1,0){.0579}}
\put(65.495,47.345){\line(1,0){.0634}}
\put(65.559,47.317){\line(1,0){.0689}}
\put(65.627,47.29){\line(1,0){.0744}}
\put(65.702,47.264){\line(1,0){.0799}}
\put(65.782,47.239){\line(1,0){.0853}}
\put(65.867,47.215){\line(1,0){.0908}}
\put(65.958,47.192){\line(1,0){.0963}}
\put(66.054,47.17){\line(1,0){.1018}}
\put(66.156,47.149){\line(1,0){.1073}}
\put(66.263,47.129){\line(1,0){.1128}}
\put(66.376,47.111){\line(1,0){.1182}}
\put(66.494,47.093){\line(1,0){.1237}}
\put(66.618,47.076){\line(1,0){.1292}}
\put(66.747,47.061){\line(1,0){.1347}}
\put(66.882,47.046){\line(1,0){.1402}}
\put(67.022,47.033){\line(1,0){.1456}}
\put(67.168,47.021){\line(1,0){.3077}}
\put(67.475,46.999){\line(1,0){.3296}}
\put(67.805,46.982){\line(1,0){.3516}}
%\end
%\qbezier(70.981,48.084)(71.873,47.08)(68.157,46.969)
\put(70.981,48.084){\line(0,-1){.0484}}
\put(71.021,48.036){\line(0,-1){.0474}}
\put(71.057,47.988){\line(0,-1){.0463}}
\put(71.086,47.942){\line(0,-1){.0453}}
\put(71.111,47.897){\line(0,-1){.0442}}
\put(71.13,47.852){\line(0,-1){.2796}}
\put(71.108,47.573){\line(0,-1){.0357}}
\put(71.083,47.537){\line(0,-1){.0346}}
\put(71.052,47.503){\line(-1,0){.036}}
\put(71.016,47.469){\line(-1,0){.0415}}
\put(70.975,47.436){\line(-1,0){.0469}}
\put(70.928,47.405){\line(-1,0){.0524}}
\put(70.876,47.375){\line(-1,0){.0579}}
\put(70.818,47.345){\line(-1,0){.0634}}
\put(70.754,47.317){\line(-1,0){.0689}}
\put(70.685,47.29){\line(-1,0){.0744}}
\put(70.611,47.264){\line(-1,0){.0798}}
\put(70.531,47.239){\line(-1,0){.0853}}
\put(70.446,47.215){\line(-1,0){.0908}}
\put(70.355,47.192){\line(-1,0){.0963}}
\put(70.259,47.17){\line(-1,0){.1018}}
\put(70.157,47.149){\line(-1,0){.1073}}
\put(70.05,47.129){\line(-1,0){.1127}}
\put(69.937,47.111){\line(-1,0){.1182}}
\put(69.819,47.093){\line(-1,0){.1237}}
\put(69.695,47.076){\line(-1,0){.1292}}
\put(69.566,47.061){\line(-1,0){.1347}}
\put(69.431,47.046){\line(-1,0){.1401}}
\put(69.291,47.033){\line(-1,0){.1456}}
\put(69.146,47.021){\line(-1,0){.3077}}
\put(68.838,46.999){\line(-1,0){.3296}}
\put(68.508,46.982){\line(-1,0){.3516}}
%\end
%\qbezier(68.157,48.864)(66.076,48.883)(65.332,48.084)
\put(68.157,48.864){\line(-1,0){.4624}}
\put(67.694,48.858){\line(-1,0){.1462}}
\put(67.548,48.851){\line(-1,0){.1423}}
\put(67.406,48.841){\line(-1,0){.1383}}
\put(67.267,48.829){\line(-1,0){.1344}}
\put(67.133,48.815){\line(-1,0){.1304}}
\put(67.003,48.798){\line(-1,0){.1265}}
\put(66.876,48.779){\line(-1,0){.1225}}
\put(66.754,48.758){\line(-1,0){.1185}}
\put(66.635,48.734){\line(-1,0){.1146}}
\put(66.52,48.707){\line(-1,0){.1106}}
\put(66.41,48.679){\line(-1,0){.1067}}
\put(66.303,48.647){\line(-1,0){.1027}}
\put(66.2,48.614){\line(-1,0){.0988}}
\put(66.102,48.578){\line(-1,0){.0948}}
\put(66.007,48.539){\line(-1,0){.0909}}
\put(65.916,48.498){\line(-1,0){.0869}}
\put(65.829,48.455){\line(-1,0){.0829}}
\put(65.746,48.409){\line(-1,0){.079}}
\put(65.667,48.361){\line(-1,0){.075}}
\put(65.592,48.311){\line(-1,0){.0711}}
\put(65.521,48.258){\line(-1,0){.0671}}
\put(65.454,48.202){\line(-1,0){.0632}}
\put(65.391,48.144){\line(0,-1){.0603}}
%\end
%\qbezier(70.498,48.53)(70.108,48.808)(68.157,48.864)
\put(70.498,48.53){\line(-1,0){.0429}}
\put(70.455,48.557){\line(-1,0){.0507}}
\put(70.404,48.583){\line(-1,0){.0585}}
\put(70.345,48.608){\line(-1,0){.0663}}
\put(70.279,48.632){\line(-1,0){.0741}}
\put(70.205,48.655){\line(-1,0){.0819}}
\put(70.123,48.677){\line(-1,0){.0897}}
\put(70.033,48.697){\line(-1,0){.0975}}
\put(69.936,48.717){\line(-1,0){.1053}}
\put(69.83,48.735){\line(-1,0){.1131}}
\put(69.717,48.753){\line(-1,0){.121}}
\put(69.596,48.769){\line(-1,0){.1288}}
\put(69.468,48.784){\line(-1,0){.1366}}
\put(69.331,48.798){\line(-1,0){.1444}}
\put(69.187,48.81){\line(-1,0){.1522}}
\put(69.035,48.822){\line(-1,0){.3278}}
\put(68.707,48.842){\line(-1,0){.359}}
\put(68.348,48.858){\line(-1,0){.1912}}
%\end
%\vector(69.16,48.827)(70.424,48.567)
\put(70.424,48.567){\vector(4,-1){.1}}\multiput(69.16,48.827)(.210667,-.043333){6}{\line(1,0){.210667}}
%\end
\end{picture}
\caption{Mach-Zehnder interferometer, in which one beam path is parallel to the rotation axis of a heavy rotating (magnetic) object $O$.}
\label{2008-flyby-f-Mach-Zehnder}
\end{figure}

Another possibility to measure possible frequency shifts of microscopic particles and light is by comparing the phases acquired along two different paths.
Consider, for the sake of demonstration, a Mach-Zehnder device depicted in Fig.~\ref{2008-flyby-f-Mach-Zehnder},
in which heavy rotating (magnetic) mass is placed along one beam path, whereas the other beam path is left unperturbed.
As usual \cite{green-horn-zei,svozil-2004-analog}, the computation proceeds by successive substitution (transition) of
states; i.e., if all components are lossless, then
\begin{equation}
\begin{array}{rcl}
S_1:\; a  &\rightarrow& ( b  +i c)/\sqrt{2}\quad , \\
O:\; b  &\rightarrow&  b e^{i \Delta \omega }\quad ,\\
S_2:\; b  &\rightarrow& ( e  + id )/\sqrt{2}\quad ,\\
S_2:\; c  &\rightarrow& ( d  + ie )/\sqrt{2}\quad .
\end{array}
\end{equation}
The resulting transition is
\begin{equation}
  a  \rightarrow \psi =i\left( {e^{i\Delta \omega} +1\over
2}\right)
d  +
\left( {e^{i\Delta \omega} -1\over 2}\right)
e  \quad .
\label{e:mz}
\end{equation}
As can be expected, for $\Delta \omega =0$ Eq.~(\ref{e:mz}) reduces to $ a  \rightarrow i d $, and an emitted quant is detected only by $D_1$.
If $\Delta \omega =\pi $, Eq.~(\ref{e:mz}) reduces to $ a  \rightarrow -  e  $, and an emitted quant is detected only by $D_2$.
If one varies the phase shift $\Delta \omega$, one obtains the following
detection probabilities:
\begin{equation}
\begin{array}{rcl}
P_{D_1}(\Delta \omega )&=&\vert ( d, \psi ) \vert^2=\cos^2({\Delta \omega \over 2})= \cos^2({{4\Omega R\over c}\;\omega}),\\
P_{D_2}(\Delta \omega )&=&\vert ( e, \psi ) \vert^2=\sin^2({\Delta \omega \over 2})=\sin^2({{4\Omega R\over c}\;\omega }).
\end{array}
\end{equation}
Taking the estimate for the hard disk configuration as before, the frequency shifts are of the order of $5 \times 10^{-7}$ times the frequency of light or of the matter waves.

Another, similar type of experiment could be done with a Michelson interferometer.
Still another setup would involve massive particles; in particular neutron interferometry \cite{rauch-werner}.

In summary, we have suggested to test the flyby of microscopic particles along heavy (magnetic) rotating objects whose rotation axis are parallel to the beam path.
Because of their sensitivity,  interferometric setups appear to be particularly promising.
The proposal might be considered as  ``unconventional'' because hard drives have not been constructed with the intention
of ``imprinting'' unidirectional magnetic macro-properties on entire disk drives, but it is not totally unreasonable to do so.
Presently it appears to be unknown whether the flyby anomaly is of gravitational and/or electromagnetic or of different \cite{adler-arXiv:0805.2895} origin --- although the heuristic formula (2)
in  Anderson {\it et al.}~\cite{anderson:091102}
does not contain any electromagnetic parameters --- thus the issue should be left open to the experiment.
In principle, one could test both the gravitational and/or electromagnetic origins with the proposed microphysical analogue.

%\bibliography{svozil}
%\bibliographystyle{osa}
%\bibliographystyle{apsrev}
%\bibliographystyle{unsrt}
%\bibliographystyle{plain}

\end{document}